\documentclass[a4paper,fleqn]{cas-dc}

\usepackage[numbers]{natbib}

\def\tsc#1{\csdef{#1}{\textsc{\lowercase{#1}}\xspace}}
\tsc{WGM}
\tsc{QE}

\begin{document}
\let\WriteBookmarks\relax
\def\floatpagepagefraction{1}
\def\textpagefraction{.001}
\let\printorcid\relax 

\shorttitle{Design and Optimization of Wearables for Human Motion Energy Harvesting}    

\shortauthors{Zhengbao Yang et al.}

\title[mode = title]{Design and Optimization of Wearables for Human Motion Energy Harvesting}

\author[1]{Weijia Peng}

\author[1]{Mingtong Chen}

\author[1]{Zhengbao Yang}
\cormark[1] 
\ead{zbyang@hk.ust} 
\ead[URL]{https://yanglab.hkust.edu.hk/}

\address[1]{The Hong Kong University of Science and Technology
Hong Kong, SAR 999077, China}

\cortext[1]{Corresponding author} 

\begin{abstract}
As wearable electronics become increasingly prevalent, there is a rise in interest and demand for sustainably designed systems that are also energy self-sufficient. The research described in this paper investigated a shoe-worn energy harvesting system designed use the mechanical energy from walking to output electrical energy. A spring is attached to electromagnetic generator embedded in the heel of the shoe to recover the vertical pressure caused by the foot strike. The simulated prototype consisted of a standard EM generator designed in MATLAB demonstrating a maximum voltage of 12V. The initial low fidelity prototype demonstrated testing the relationship between the EM generator and a simple electrical circuit, with energy output observed. Future research will explore enhancing the overall generator design, integrate a power management IC for battery protect and regulation, and combine the system into a final product, wearable footwear. This research lays a foundation for self-powered footwear and energy independent wearable electronic devices.

\end{abstract}




\maketitle

\section{Introduction}

Nowadays wearable electronics are becoming a common thing in everyday lives. They are usually low power devices for functions such as health monitoring, fitness tracking, and smart interactions. Most of the devices still rely heavily on batteries and lead to frequent recharging. As a result, the development of sustainable and self-chargeable systems has brought significant interest in recent years.

Human motion such as walking, waving, running offers a consistent source of mechanical energy that can be harvested to power wearable devices. Take walking and running as an example, every step can generate significant mechanical force, which makes shoes an important platform for combining energy harvesting system and low power devices. By converting the kinetic energy from walking into electrical energy, a self-sustaining shoe with low power electronics and no need to charge becomes possible. 

This leads to a compelling application on the development of smart footwear. For example, Nike has announced a smart shoe named “Adapt BB,” which utilizes and electrical self-lacing mechanism that make users easily tie their shoes without even touching the shoes. However, that shoes depend on traditional battery charging, which increases user burden. This project explores the feasibility of replacing the power supply with an energy harvesting system that can store electricity directly from walking in daily life and eliminate the need for charging. 

As low-power electronics and smart wearable technologies increase, harvesting energy from human motion has emerged as a sustainable option to power devices without traditional batteries. Among a variety of human motion sources, walking provides the most consistent and biomechanical energetic behavior, which makes footwear a natural substrate for biomechanical energy harvesting.

The heel pad-based assistance (HPA) device introduced by Pan et al. \cite{1} not only harvested energy from walking but assisted gait while performing an important supportive function during the gait cycle for shock absorption. The system uniquely activated collisional heel-strike energy into electricity using a unique mechanism combining a spring-loaded, gear-reduced motion that linked to a rotary electromagnetic generator. The device prototype demonstrated a peak power of 3.8 ± 0.3 W, reduced muscle activation utilization, and improved walking economy resulting in approximately 10\% less metabolic cost during normal alternatives walking. This device demonstrates the possibilities of seamless, multi-functional systems providing both assistive function and wearable power applications where energy harvesting is coordinated into a heel-embedded energy harvesting system.

Yun et al.\cite{2} developed an exo-shoe triboelectric nanogenerator (ES-TENG) with a bi-directional gearbox to translate low-frequency stepping motion into the high-speed rotation movement of a rotational TENG to focus on power generation. ES-TENG developed a specific power of up to 13 µW/g, while concurrently attaining peak open-circuit voltage greater than 3 kV, illustrating its efficiency of power generation, additionally, potential therapeutic or stimulation application. This gearbox-based transformation provides coherent energy harvesting ability during compressing and releasing phase of gait, which would be beneficial on wearable devices and energy harvesting.

Yin et al. \cite{3} describe a shoe mounted piezoelectric energy harvester (PEH) construction using a frequency up-conversion mechanism which uniquely employs ratchets and piezoelectric cantilever beams together. The construction generates high-frequency oscillations through the low-frequency gait input yielding a peak power of 13.88 mW (upwards phase), while when it balances in the downwards phase generating a stable output around 8 mW. The system provided consistent results with gait at a walking frequency ranging from 1.5-4 Hz which varies with speed and subject behavior. The introduction of the potential of structural tuning and processing mechanical amplification is important to understand future output behavior of piezoelectric harvesters.

The objective of this project is to design an energy harvesting device in a size that can be embedded into a shoe. Utilizing an electromagnetic generator to convert the mechanical energy of walking into electrical energy. The energy will be stored inside the shoe and drive some low power designed devices such as a self-lacing mechanism, make the shoe able to have some function without the use of external power sources. This report will present a detailed overview of the design process, from the theoretical background and system architecture to the practical implementation and performance evaluation of the proposed energy harvesting footwear system.

\section{Design and Methodology}

\begin{figure}[h]
	\centering
		\includegraphics[scale=0.6]{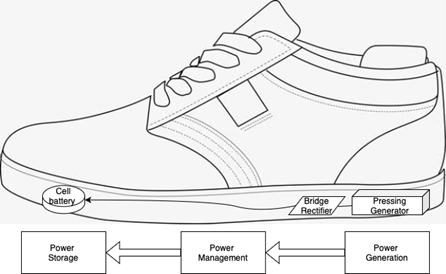}
	  \caption{Overall System Schematic}\label{FIG:1}
\end{figure}

\begin{figure}[h]
	\centering
		\includegraphics[scale=1]{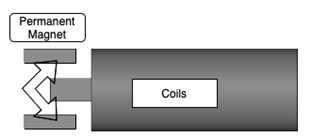}
	  \caption{Generator Schematic}\label{FIG:2}
\end{figure}

\begin{figure*}[h]
	\centering
		\includegraphics[scale=1]{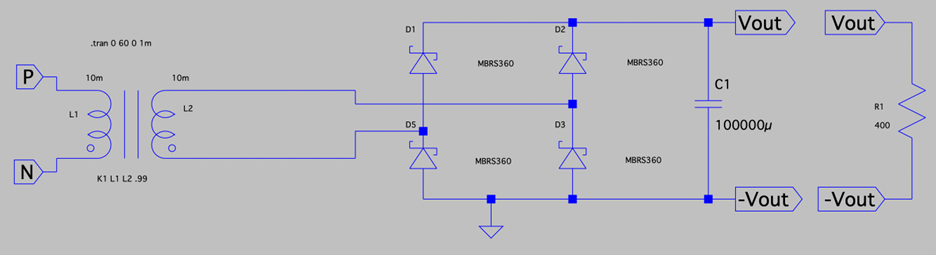}
	  \caption{Bridge Rectifier Schematic}\label{FIG:3}
\end{figure*}

The goal of this project is to develop a small, energy harvesting system embedded in a shoe that converts the mechanical energy from walking into electrical energy, enough to power a low-power actuator, such as a self-lacing system. A morphological chart (morph chart) was used to inform several of the surrounding designs, functioning areas of the shoe, for the concept development.

By comparing the resulting concepts based on feasibility of the concept developed, energy to power produced, complexity of integration, and overall comfort for the user, the magnet-in-coil oscillating electromagnetic generator internally integrated, powered by compression through the foot, storing energy in a Li-ion coin cell, and managed by energy harvesting IC, was selected for further development. The selected concept has a good balance of user comfort, durability and easy to integrate. 

The final design will consist of a small, compact spring-loaded micro electromagnetic generator module integrated in the shoe insole or heel area. As the user walks, compressing the foot creates pressure within the micro generator module which produces alternating current through movement of a magnet and coil. The alternating current is converted and regulated to direct current through a low-power harvesting IC, and the output is stored in a rechargeable coin cell battery. This stored energy can be used to power some low-power devices. This method takes advantage of actual gait biomechanics, along with accessible materials, to create a wearable power source with promising future possibilities for adoption into smart shoes and other IoT wearables.

\begin{table*}[h]
\caption{Morphological Chart for Wearable Energy Harvesting Shoe System}\label{tbl1}
\begin{tabular*}{\tblwidth}{lcccc}
\toprule
 Function & Idea 1 & Idea 2& Idea 3 &Idea 4 \\ 
\midrule
 Power Generation &Piezoelectric&Linear Electromagnetic&Rotary Electromagnetic& Magnet-in-Coil Oscillating \\
 Installation Format &External Unattachable& External Attachable & Internal & / \\ 
 Energy Source & Heel Strike& Foot Motion Inertia & Full Foot Compression & / \\ 
 Energy Storage & Super Capacitor& Li-ion Coin Cell & Thin-film Flexible Battery & / \\  
 Power Management & Bridge rectifier & Energy Harvesting IC & / & / \\ 
\bottomrule
\end{tabular*}
\end{table*}

Figure 1 below shows the detailed schematic of the whole system. The system contains three parts: Power Generation, Power Management, and Power Storage. As described in previous section, a power generator applying the principle of “Magnet-in-Coil Oscillating” will be used as the power generation part. However, since the generator will only generate alternative current, a power management part with the ability to convert AC voltage into DC is necessary in the next part. For the power storage part, since the space is very limit inside a shoe, power storage device with the smaller size is the better choice. A Li-ion coin cell battery fits the situation perfectly since it is both rechargeable and has a small, flat size that can fit inside a shoe, with a high durability.

The micro press generator in this project operates in accordance with electromagnetic induction via Faraday's Law. Faraday's Law states that the change in magnetic flux through a coil creates an electromotive force (emf) across the coil. In this case, footfalls which in the form of mechanical energy, is converted into electrical energy by utilizing the change in magnetic field caused by the shock made during the contact of foot and floor. The electrical energy is generated by mainly three components: Permanent Magnet, Coil, and the moving motion of coil, shown as the schematic below in figure 2.

The magnet inside the coil will attach to one of the permanent magnets on the top or bottom, when external force is applied and change the magnet it attaches to, the magnetic field inside the coil changes, thus electricity is generated. In the project’s design, spring is added to the generator to ensure the rebound of coil after each compression, corresponding to the steps during walking. Which bring up the following stages: Neutral Stage:  When there is no external force acting on the generator, the spring will carry the equity onto the upper magnet and the coil, demonstrating little magnetic flux linkage. Compression Stage: When users’ foot fall on ground, there is a compressive force apply to the generator, causing the connected magnet to change from the upper to lower, and a change in the magnetic flux. Induction Stage: When the magnetic flux changes, an emf in the coil causes alternating current and a voltage out, which is the electrical energy converted by kinetic energy. Recovery Stage: After the user releases the footsteps, the spring will restore the generator back onto its original position and be prepared for the next cycle.

The electrical energy generated by the generator is in the form of alternative current (AC), where the battery needs to be charged with direct current (DC). So, the post process of energy generated by the generator is important. To convert AC voltage into DC voltage, the Bridge Rectifier would be a great option. Considering the low voltage generated, bridge rectifier utilizing Schottky diode followed by a boost converter is a reasonable choice. 
The rectified output is still relatively low voltage, typically in the range of 2–5 V, yet is likely to be variable with step force and frequency. Additionally, a smoothing capacitor is applied at the rectifier output to smooth out any voltage ripple and to store energy temporarily for the next time the footstep induces the energy storage.

Rechargeable Li-ion coin cell was chosen for the harvested energy storage due to its small size and rechargeable characteristics. The storage unit allows the system to discharging the stored energy to provide power to the low-power devices such as a self-lacing mechanism.

\section{Simulation and Validation}

Simulations are made through MATLAB and LTspice to validate the design. Making sure that the electrical energy generated by the generator can be rectify by the bridge rectifier for charging the coin cell battery at the end of the system. MATLAB was used to simulate the performance of the generator and LTspice was used to simulate the rectification and voltage buffering process. In MATLAB, the behavior of the electromagnetic generator was modeled based on Faraday’s Law, which states that the voltage in proportional to the time derivative of the magnetic flux through the coil. The magnetic field was first assumed, then induced electromotive force was calculated using the Faraday’s Law.The result shows a peak voltage about 12.1366 V.

In LTspice, a full-wave Schottky bridge rectifier circuit was simulated with the MBRS360 Diodes. The Schottky diodes were chosen due to their low forward voltage drop, fast switching speed, and high efficiency at low voltages. The generator was modeled as a voltage source with pulse signal. The value of 10V was assumed based on the MATLAB simulation result. The following graph shows the schematic of the circuit designed in LTspice. A capacitor is added at the end to smooth the voltage.

The results of LTspice simulation shows that the circuit has the ability to stable and convert the AC voltage from the generator to DC voltage that can be used to charge the battery. However, as the result below shows, it needs a starting stage for the voltage to be stable at the designated voltage. In this simulation, a more realistic frequency of walking was assumed to be 1 step per second, at around 80 seconds, the voltage tends to become stable.

\begin{figure}[h]
	\centering
		\includegraphics[scale=0.8]{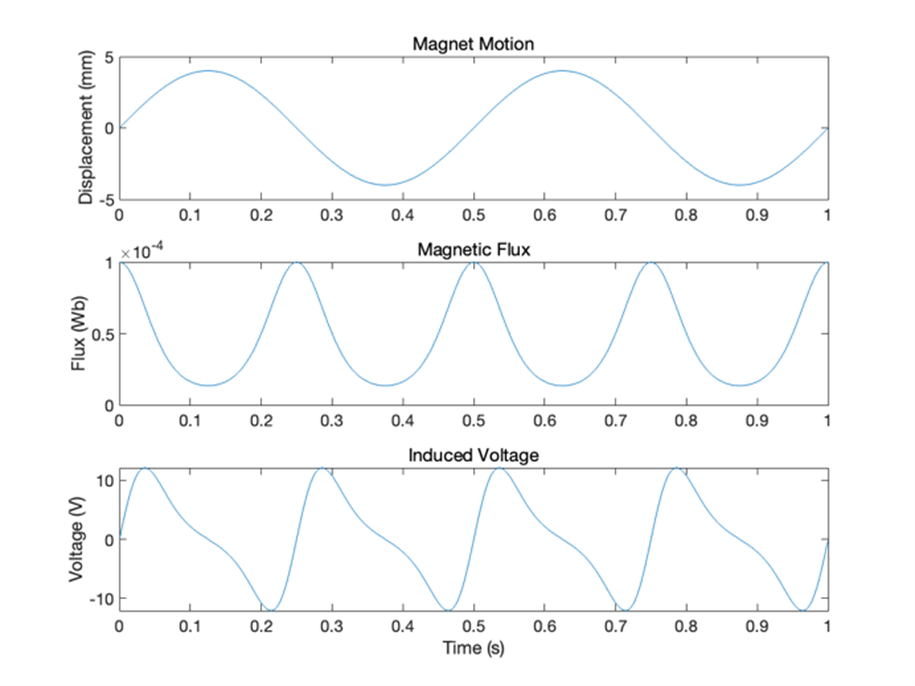}
	  \caption{MATLAB Simulation results}\label{FIG:4}
\end{figure}

\begin{figure*}[h]
	\centering
		\includegraphics[scale=1]{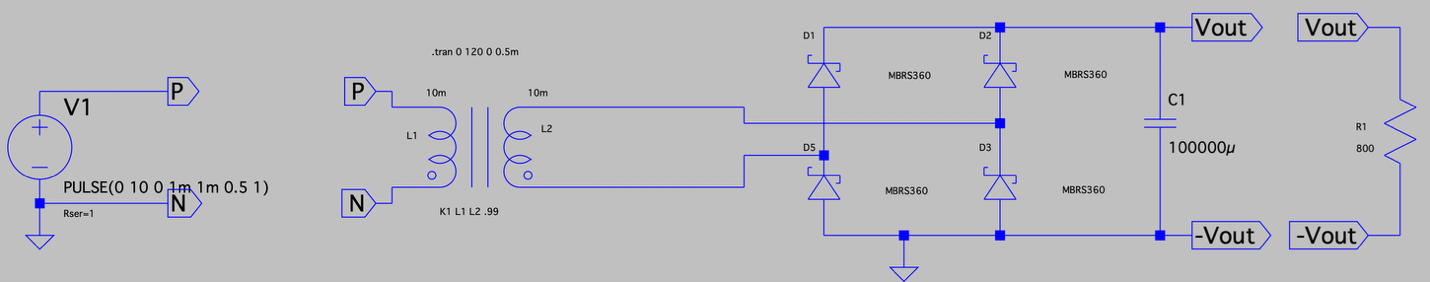}
	  \caption{LTspice Schematic}\label{FIG:5}
\end{figure*}

\begin{figure}[h]
	\centering
		\includegraphics[scale=0.7]{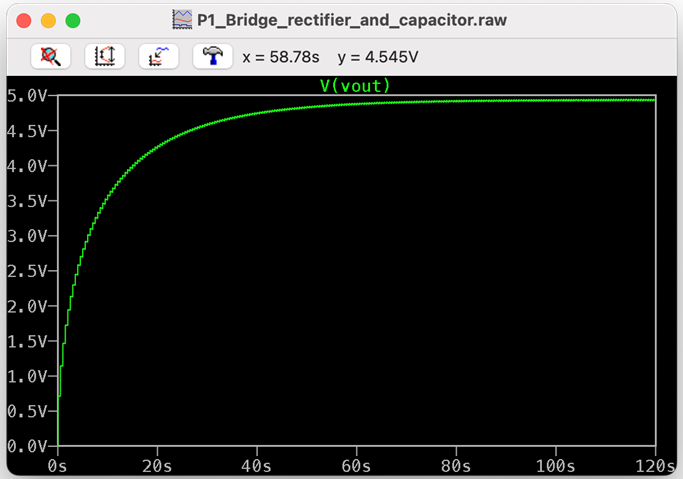}
	  \caption{LTspice simulation results}\label{FIG:6}
\end{figure}

\begin{figure*}[h]
	\centering
		\includegraphics[scale=1]{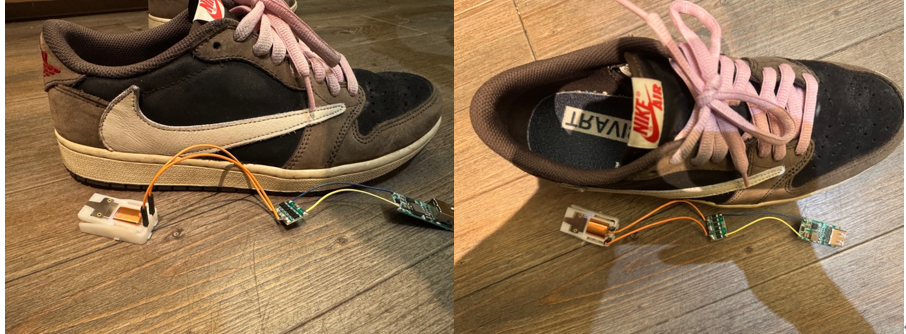}
	  \caption{Low Fidelity Prototype with Comparison to Actual Shoe}\label{FIG:7}
\end{figure*}

\begin{figure*}[h]
	\centering
		\includegraphics[scale=1]{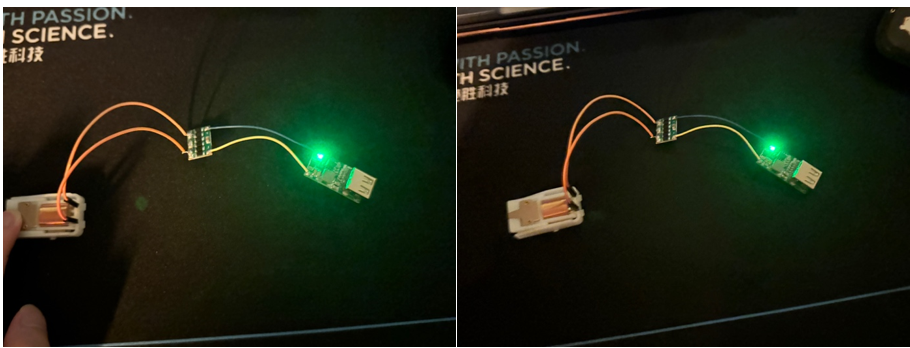}
	  \caption{LED Indicator Under Just-Pressed and Just-Released Condition}\label{FIG:8}
\end{figure*}

Only a low-fidelity prototype was developed for this project due to time constrains. The main purpose of developing the prototype is to validate essential functions of the system. The focus of the prototype is to demonstrate a generator that can convert the kinetic energy by walking to electrical energy with the electromagnetic generator. 
The CAD model of the generator was developed through SolidWorks. A housing was designed to fit a spring that allows it to bounce back after pressing, which is corresponding to the situation of stepping on the ground. When the user’s foot leaves the ground and prepares for next contact to ground, the position of magnet restores to its original state due to the spring. The CAD model provides a better way to visualize the generator design. 

For the bridge rectifier part of the system, a low fidelity prototype was planned to make with breadboard with diodes, then connect with the generator. However, since the components attach to breadboard connection is loose, the first low fidelity prototype of the circuit is lost. The new method of building a low fidelity is to use soldering technique to connect a prebuild bridge rectifier and the coin cell battery. The following graph shows the current low fidelity prototype. It meets the requirements of having a small size that can fit in a shoe.

In the low fidelity prototype, the power storage part was replaced with a USB output port with LED indicator. The reason of doing this is to have a clearer visualization of how the alternative current generated by the generator converted to direct current that can power or charge other devices. As figure 8 shown, the LED indicator did light on for a short time when the generator was pressed and restore to original state by the spring. Showing that the bridge rectifier successfully converts the output voltage in to DC voltage.

\section{Discussion and Conclusion}

At this point, the project remains in its prototype phase, consisting of a low-fidelity generator module assembled and some basic electrical circuitry tested to confirm basic connection possibility. Although the first assembly has helped validate the concept fundamentally, the prototype only has limited task capability and does not currently demonstrate functional energy harvesting or delivered energy output, therefore, many improvements and development steps have been identified to move toward a functional integrated energy-harvesting shoe project.
The next step will be to improve the mechanical design of the generator to optimize energy conversion efficiency. This includes adjusting the alignment between the magnet and coil, enhancing the stiffness of the internal spring, and creating repeatable and smooth motion, with the most basic adjustments being walking. The new generator design will also consider better mechanical integration into the shoe sole to encompass user physical comfort and longevity.

Developing a complete mid-fidelity prototype will allow for the first testing portion of a shoe wear environment and include the improvements made to the generator, and the optimized circuit integrated with energy storage. Testing will focus primarily on the number of steps or amount of foot movement required to accrue energy and acquire usable energy, whether enough energy can activate another load, and user comfort while walking.
The above steps will move the project forward from a conceptual prototype to a demonstrable energy harvesting wearable device that will be able to sustainably and independently power shoe-based low-energy electronics.

This project considered the design, simulation, and low-fidelity prototyping of a wearable energy harvesting system into footwear. The general concept was to generate electrical energy from the mechanical energy of walking using a miniaturized electromagnetic generator, then store it in a Li-ion coin cell, which can then drive low-power electronics, like a self-lacing shoe mechanism. 
The project scoped the literature to find its position in the state-of-the-art of wearable energy harvesting, including many advances, such as triboelectric nanogenerators and piezoelectric systems. Ultimately, electromagnetic generators were chosen for their durability, suitability for the compression driven by footsteps, and the relatively straightforward nature of designing an electrical interface.

The design process began with a morphological chart to compare functional concepts across generator types, types of displacements, power management, and power storage. The final concept utilizes a horizontally oriented, magnet-in-coil electromagnetic generator, placed inside the heel, a bridge rectifier with Schottky diodes, and a rechargeable Li-ion coin cell as energy storage.
MATLAB simulations showed that the generator may theoretically allow the production of upwards of 12 V of peak voltage for each compression event, lending credibility to the energy potential of the proposed system. LTspice simulated the rectification circuit and produced usable DC voltages from AC pulses produced from the electromagnetic generator. It was shown that due to the smoothing capacitor, the pulsed wave was converted and stabilized into usable DC voltage. The LTspice simulations show that the system architecture was valid.
CAD of the generator was made to provide visualizations of the detailed design for the power generation part of the whole system. A low-fidelity prototype was constructed to engage with mechanical design, component sizing, and basic electrical connection. While not fully functional, the low-fidelity prototype tested and validated the core concepts of assembly. 
While this project made significant progress, more refinements can be made. The power management section of the system can be optimized with extra components such as power ICs and customized PCBs. Nonetheless, this project provided a healthy basis, demonstrated that this concept could work, and illustrated a pathway toward a fully integrated mid-fidelity prototype.In summary, this project was successful for the designing, simulating, and beginning the prototyping of a wearable energy harvesting system. With a few more adjustments and the addition of a power management IC, this system may evolve a self-powered smart shoe platform used for autonomously powering low-energy applications during walking. This could serve as a leap toward a more practical and sustainable wearable electronic.










\bibliographystyle{cas-model2-names}

\bibliography{cas-refs}



\end{document}